\documentclass[11pt,a4paper]{article}
\usepackage{jinstpub}

\usepackage{pdfpages,multirow,ragged2e}
\usepackage{svg}
\usepackage{wrapfig}
\usepackage{listings}
\usepackage[utf8]{inputenc}
\usepackage[USenglish]{babel}
\usepackage[caption=false]{subfig}

\graphicspath{ {./Figures/} }

\title{Machine Learning for Reducing Noise in\\RF Control Signals at Industrial Accelerators}

\author[a,1]{M. Henderson, \note{Corresponding author.}}
\author[a]{J. P. Edelen, }
\author[a]{J. Einstein-Curtis, }
\author[a]{C. C. Hall,}
\author[b]{\\J. A. Diaz Cruz, }
\author[b]{and A. L. Edelen}

\affiliation[a]{RadiaSoft LLC, Boulder CO, USA}
\affiliation[b]{SLAC, Menlo Park CA, USA}

\emailAdd{mhenderson@radiasoft.net}

\abstract{Industrial particle accelerators typically operate in dirtier environments than research accelerators, leading to increased noise in RF and electronic systems. Furthermore, given that industrial accelerators are mass produced, less attention is given to optimizing the performance of individual systems. As a result, industrial accelerators tend to underperform their own hardware capabilities. Improving signal processing for these machines will improve cost and time margins for deployment, helping to meet the growing demand for accelerators for medical sterilization, food irradiation, cancer treatment, and imaging. Our work focuses on using machine learning techniques to reduce noise in RF signals used for pulse-to-pulse feedback in industrial accelerators. Here we review our algorithms and observed results for simulated RF systems, and discuss next steps with the ultimate goal of deployment on industrial systems.}


\begin{document}
\maketitle

\section{Introduction}

Machine learning (ML) has been identified as having the potential for significant impact on the modeling, operation, and control of particle accelerators~\cite{edelen2016, edelen2018}. For machine diagnostics specifically, there have been numerous efforts to improve measurement capabilities and detect faulty instruments. For example, many developments have focused on improving beam position monitors (BPMs), including the removal of poorly performing BPMs. Work done at the Large Hadron Collider (LHC) identified faulty BPMs prior to application of standard optics correction algorithms~\cite{fol2019}. More recently, ML methods have improved optics measurements from beam position monitor data~\cite{fol2018}. 

While ML continues to be a popular area of research for accelerator diagnostics, there is a lack of engineering knowledge when it comes to the application of ML for RF systems. As the demands on industrial accelerators increase, so does their complexity and the need for refined control methods. While ML techniques have the potential to improve accelerator operations generally, they show particular promise for systems operating in industrial environments. The ability to improve signal to noise ratio and extract key characteristics from RF signals would greatly improve the ability of industrial systems to meet growing performance demands.  

Autoencoders (AEs) are an ML technique with established application to noise removal for diagnostic signals. Variational Autoencoders (VAEs) are especially adept at removing noise due to the enforcement of a smoothness criterion in the model latent-space \cite{higgins2017}. This feature of VAEs has seen them be applied to several physical process models, such as those in gravitational wave research~\cite{ormiston2020,vajente2020} and geophysical data~\cite{bhowick2019}. Variational recurrent autoencoders (VRAEs) have the added advantage of being well-suited to sequential data. Autoencoders have also been applied to simulated ring BPM data to remove both additive white Gaussian (AWG) noise and noise of different colors (i.e., power law spectra)~\cite{edelen2022}.

In this paper, we evaluate several ML methods for noise removal, including AEs, convolutional AEs (CAEs), and VRAEs, and compare them with a more conventional Kalman filter (KF) approach. We begin with a review of our data generation pipeline, followed by descriptions of our models and their individual performance in the removal of noise from RF signals. Finally, we compare these results and evaluate the performance of a CAE for noise removal in measurement data collected on an industrial RF system. 

\section{Data Generation}
Data were generated using an RF simulator created by RadiaSoft for developing RF control algorithms, IOC software, and user interfaces. The simulator is integrated with EPICS and can be run through a number of modalities, including command line interfaces, Python scripts, Jupyter notebooks, and direct EPICS connections. It describes RF signal propagation using a standard scattering matrix approach and cavity dynamics using an effective RLC circuit model which accounts for a number of system characteristics, including: coupling and quality factors, frequency, drive amplitude and phase, pulse duration, and cavity detuning. Model dynamics are based on equations derived in~\cite{czarski2006, echevarria2018a, echevarria2018b}, and additive white Gaussian noise (AWGN) is applied to measurements.

Diverse training and testing datasets were generated by varying RF pulse and cavity characteristics. The cavity resonance frequency was chosen to be 2856 MHz, a typical value for industrial applications. Pulse lengths were varied from 3\textendash7.5 $\mu$s, a reasonable range for industrial accelerator applications operating at S-Band. Additionally, we varied RF pulse amplitudes and start times within the data window. Although pulses don't typically vary in position along DAQ windows, this flexibility encourages better generalization when transferring from simulations to measurements.

The RF cavity parameters of interest for this study were the quality $Q$ and coupling $\beta$ factors, which were varied over ranges of $Q=1.00\times10^{4}$~\textendash~$2.25\times10^5$ and $\beta=1$~\textendash~$3$. Detuning ($\Delta\omega=\omega_0-\omega_D$) was also varied within a range of one half bandwidth above or below cavity resonance, a typical range in industrial systems. Overall, the parameters chosen represent reasonable values for industrial RF systems and allowed us to develop simulation-based algorithms that should transfer well to actual measurements.

\section{Overview of Methods}

\subsection{Kalman Filter}
The standard Kalman filter~\cite{kalman1960} (KF) is a class of linear quadratic estimators which feature a predictor-corrector update scheme. In the prediction step, a dynamical model is used to project the current state of a system forward in time. This is followed by a corrector step for assimilating incoming data in which Bayesian statistics are used to evaluate the likelihood of a measurement given the current best dynamical estimate and update the estimate accordingly.

The state in our KF model consists of input RF signals along with transmitted and reflected cavity signals. State transition dynamics are given by the linear differential equation in our equivalent circuit model. We chose a measurement model that mimics read-back data typically available in applied settings. In this context, we take KF mean estimates as predictions of the true (noiseless) state and estimate error covariances as uncertainty metrics.

\subsection{Feed-forward Autoencoder}

The first of our ML methods for noise reduction is an AE based on feed-forward neural networks (FNNs). Individual RF signals are used as inputs to this model such that each time-step is a feature in the input space. For this study we use one model trained on all waveform types (forward, reflected, and transmitted). The architecture of this model is relatively simple, with a single encoder layer and a latent dimension of $32$. The model was trained using noisy waveform data from our simulator in an unsupervised fashion, with inputs and target outputs being the same noisy signal. In this context, the mechanism for noise reduction is the inability of the latent space representation to encode noise due to the lack of structured information present in the data. 

\subsection{Convolutional Autoencoder}

Convolutional neural networks (CNNs) are adept at feature extraction, especially for data with translationally invariant features arising from underlying contexts or dynamics. Our CAE models leverage this capability to extract latent information from noisy RF signal input sequences which are then decoded to reconstruct noiseless signals. As with the feed-forward AE, the lack of predictability in the noise leads to reconstructions which preferentially follow the true, noiseless dynamics.

Our CAE model architecture consists of a sequence of 1-D convolutional and max-pooling layers that reduce the temporal dimension of the signals to create an increasingly dense feature space representation. This results in the compression of input data into a latent space representation composed of of 30 variables. We then used up-sampling and convolutional layers to ultimately reconstruct the waveform. This structure is very similar to that of the U-net~\cite{ronneberger2015}, a model often used for image segmentation and other machine vision problems, but without skip connections (which in this case would only trivialize the decoding task of the network).

 When training CAE models, we treat each waveform as a unique input to allow the CAE to learn noise rejection regardless of the type of data being processed (i.e., forward, reflected, or probe signals). Along with the signal translations described above, this serves to improve the generalizability of the CAE when considering data collected on different types of machines where probes are not always available or traveling wave structures where the signal envelopes do not follow the same profile as standing wave cavities.

\subsection{Variational Recurrent Autoencoders}

In contrast to standard AE architectures, the encoder in a VAE maps inputs to a latent space distribution rather than a single latent vector. The latent distribution is parameterized by a mean vector and diagonal covariance matrix, and is sampled during training to produce the vectors which are passed through the decoder to create reconstructions. The loss function for a VAE contains an extra term, the Kullback–Leibler (KL) divergence~\cite{kingma2014}, in addition to the reconstruction loss used by most AEs. This term acts as a regularization metric and enforces smoothness in the learned latent space. This training scheme helps prevent overfitting and can produce a more interpretable latent spaces, which makes VAEs excellent tools for data generation and noise elimination.

Our VAE employs recurrent NNs (RNNs) for the encoder and decoder networks to better capture temporal relationships between input data, distinguishing it as a VRAE~\cite{fabius2015}. The RNNs used in our models are long short-term memory (LSTM) units, and latent space distributions are taken to be Gaussian. As the noise is also known to be Gaussian-distributed, reconstruction loss is taken as the mean squared error between inputs and reconstructions.

\subsection{Noise Reduction Analysis}

To evaluate the noise reduction capabilities of our models, we compute residuals between model reconstructions $\hat{X}$ and the noiseless ground-truth signals $X$:

\begin{equation}
\label{eq:residuals}
N(t)=\hat{X}(t)-X(t)
\end{equation}

To provide salient comparisons between residuals and the AWGN profiles of noisy ground-truth data, we then compute the residual power spectral density (PSD) via fast Fourier transform:

\begin{equation}
\label{eq:PSD}
P_N(\omega) = \left|\ \int N(t)\ e^{-2\pi i \omega t} dt\ \right|
\end{equation}

\noindent Finally, we integrate the PSD over the frequency domain to compute total integrated noise:

\begin{equation}
\label{eq:int-noise}
N_\mathrm{int} = \int P_N(\omega)\ d\omega
\end{equation}

We use both the PSD and the integrated noise to perform comparisons between each of our models as well as with the known original noise profile. Note that the latter is only possible in our study as we are using simulated data with an additive noise model, whereas such comparisons would not be possible in a real industrial RF system.

\section{Results}

\subsection{Single Sample Analysis}

\begin{figure}[h!]
  \begin{minipage}[b]{0.475\textwidth}
    \centering
    \includegraphics[width=.85\linewidth]{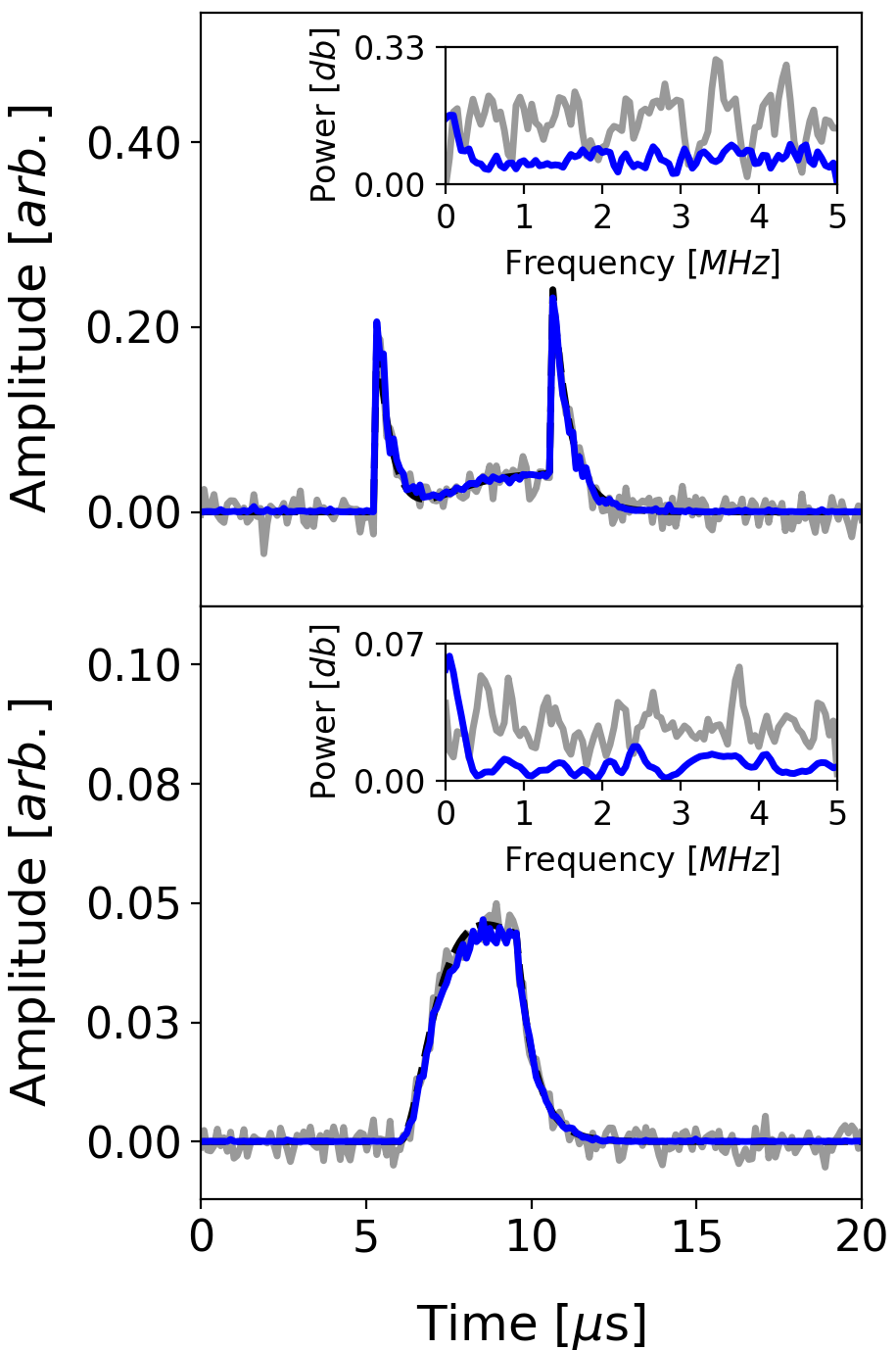}
    \caption{Average (top) and best case (bottom) reconstructions and PSDs (inset) for the KF model.}
    \label{fig:kf_performance}
  \end{minipage}
  \hfill
  \begin{minipage}[b]{0.475\textwidth}
    \centering
    \includegraphics[width=.85\textwidth]{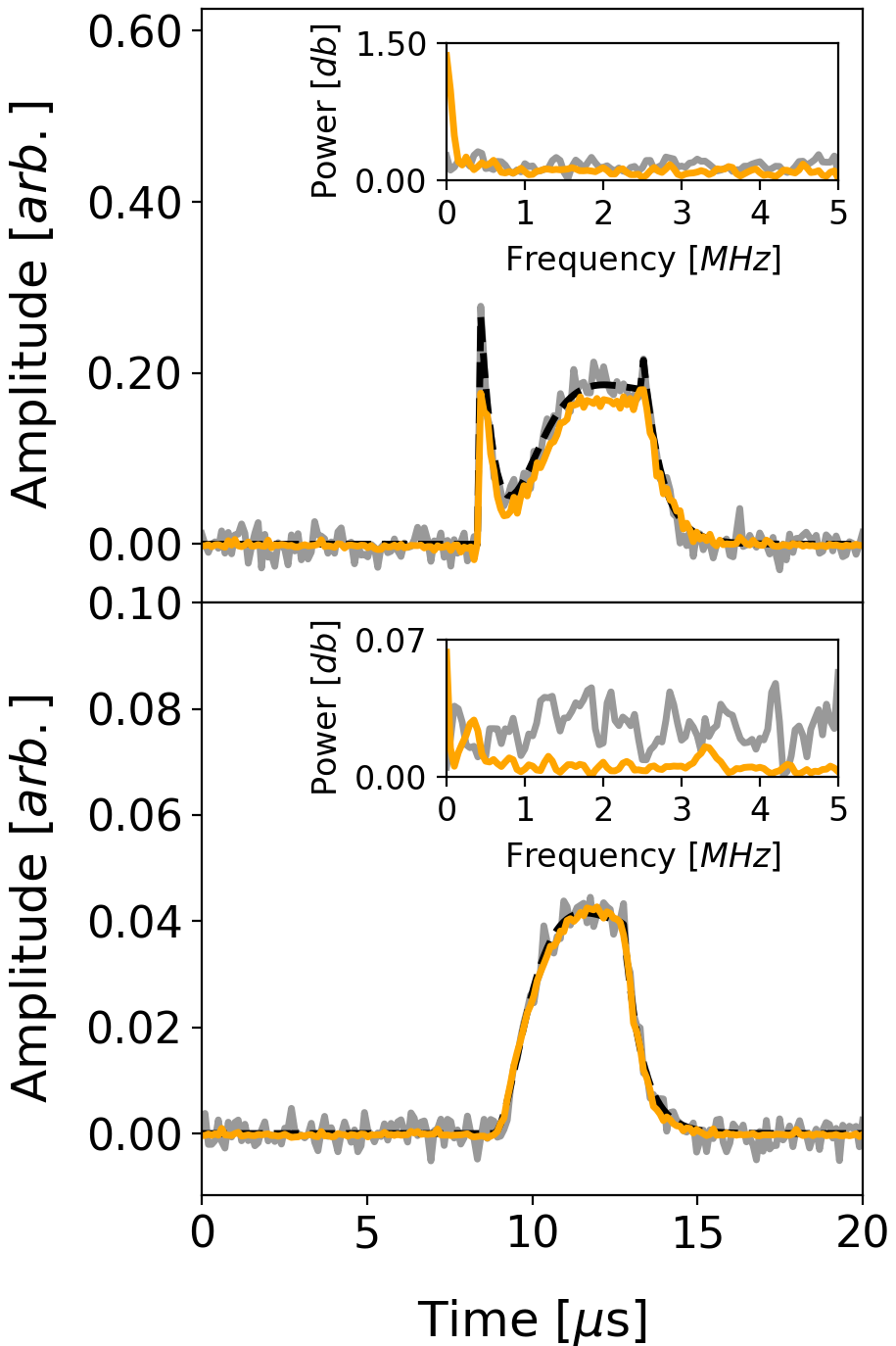}
    \caption{Average (top) and best case (bottom) reconstructions and PSDs (inset) for the AE model.}
    \label{fig:ae_performance}
  \end{minipage}
\end{figure}

\begin{figure}[h!]
  \begin{minipage}[b]{0.475\textwidth}
    \centering
    \includegraphics[width=.85\linewidth]{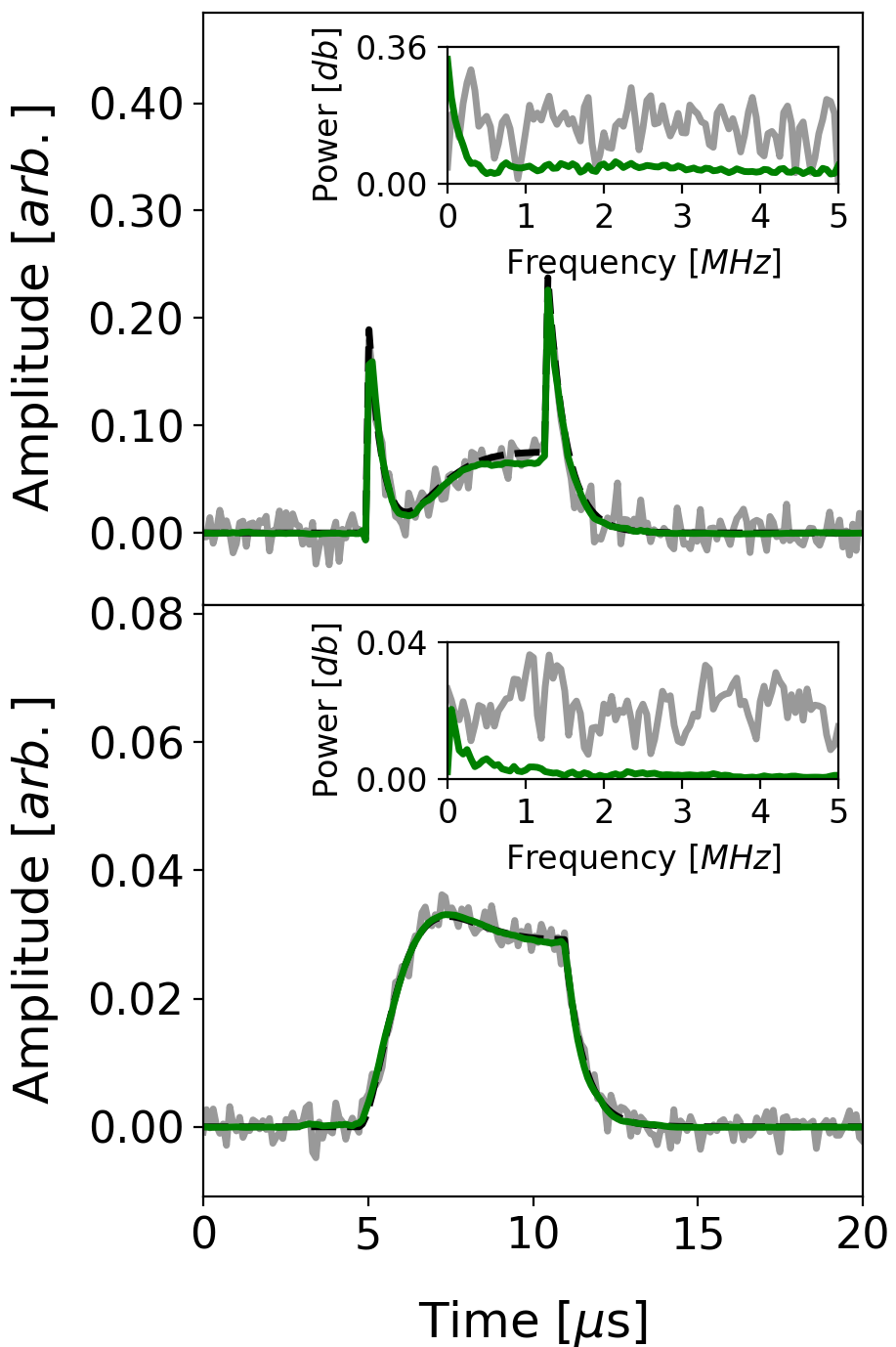}
    \caption{Average (top) and best case (bottom) reconstructions and PSDs (inset) for the CAE model.}
    \label{fig:cae_performance}
  \end{minipage}
  \hfill
  \begin{minipage}[b]{0.475\textwidth}
    \centering
    \includegraphics[width=.85\textwidth]{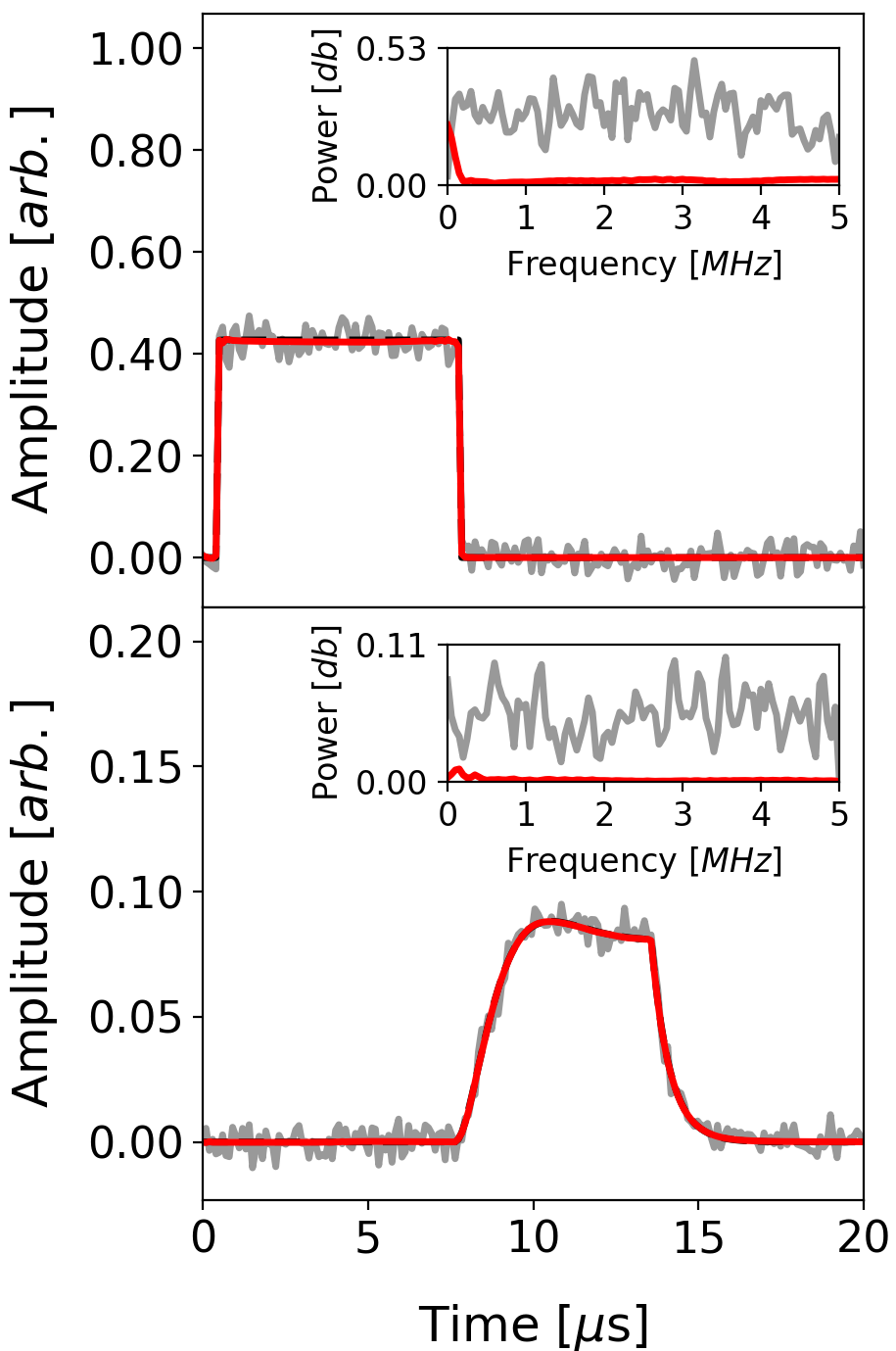}
    \caption{Average (top) and best case (bottom) reconstructions and PSDs (inset) for the VRAE model.}
    \label{fig:vrae_performance}
  \end{minipage}
\end{figure}

Each of our models succeeded at reducing noise in a test set of RF waveform data. Figures \ref{fig:kf_performance}-\ref{fig:vrae_performance} demonstrate the results for each model on single waveform samples for which the model achieved its median and maximum performance, as dictated by the total integrated noise (Eq. \ref{eq:int-noise}).

\paragraph{Kalman Filter}

The KF model successfully reconstructed RF signals with reduced noise across the full spectrum when compared with the AWGN baseline. However, for both the average and best test case (see Fig. \ref{fig:kf_performance}), the level of residual noise is \textit{just below} the baseline. This indicates only moderate noise reduction capabilities which are not unexpected given the reliance of the KF on a dynamical model and noisy measurements for contextual information about the system.

\paragraph{Feedforward Autoencoder}

Unlike the KF, the average test case for the feed-forward AE model resulted in a PSD which approximately equals the AWGN baseline across the frequency domain (see Fig. \ref{fig:ae_performance}). The best-case performance, however, resulted in a near-zero PSD at frequencies above 0.5 MHz which was equal to or lower than that observed in the best-case for the KF. Furthermore, the AE was found to perform consistently better for transmitted and reflected signals (average MSE $3\times10^{-4}$ and $9\times10^{-6}$, respectively), and worse for the forward signal (average MSE $1
\times10^{-3}$). This indicates that the AE model is in general capable of extracting sufficiently rich contextual information from waveform data to achieve high-fidelity reconstructions, but that the learned representation is not robust enough to generalize.

\paragraph{Convolutional Autoencoder}

The CAE model provides a much better demonstration of noise reduction than either the KF or the AE models (see Fig. \ref{fig:cae_performance}). Similar to the best-case performance of the AE, the \textit{average} CAE test case shows a near-zero PSD at frequencies above 1 MHz and the best case exhibits even lower PSD values beginning at around 0.5 MHz, with a noticeable peak in the PSD at DC. This performance reflects the improved ability of convolutional models to efficiently extract contextual information from inputs when compared to feedforward network models. The relative uniformity of CAE performance for a variety of waveform profile and pulse start time and duration is indicative of the translational invariance of the patterns identified by the model in the 1D sequences used during training.

\paragraph{Variational Recurrent Autoencoder}

The VRAE model was found to produce the highest fidelity reconstructions of noiseless signals given noisy inputs (see Fig. \ref{fig:vrae_performance}). The PSD for the VRAE reaches near-zero values well below a frequency of 0.5 MHz in even the average test case, and is near-zero across the spectrum in the best case. The consistency of this performance indicates a strong ability to extract contextual information from input signals not dissimilar to the CAE, but in this case provided by the sequence-awareness of the underlying LSTM networks employed by the VRAE. Although the VRAE lacks the convolutional mechanisms of the CAE which extract translationally invariant patterns, the smoothness of the learned latent space promoted by the KL divergence loss term nonetheless supports very robust performance across all waveform profiles, pulse start times and durations, and system parameters like cavity coupling.

\subsection{Statistical Comparisons}

\begin{wrapfigure}{r}{0.475\textwidth}
    \centering
    \includegraphics[width=.9\linewidth] {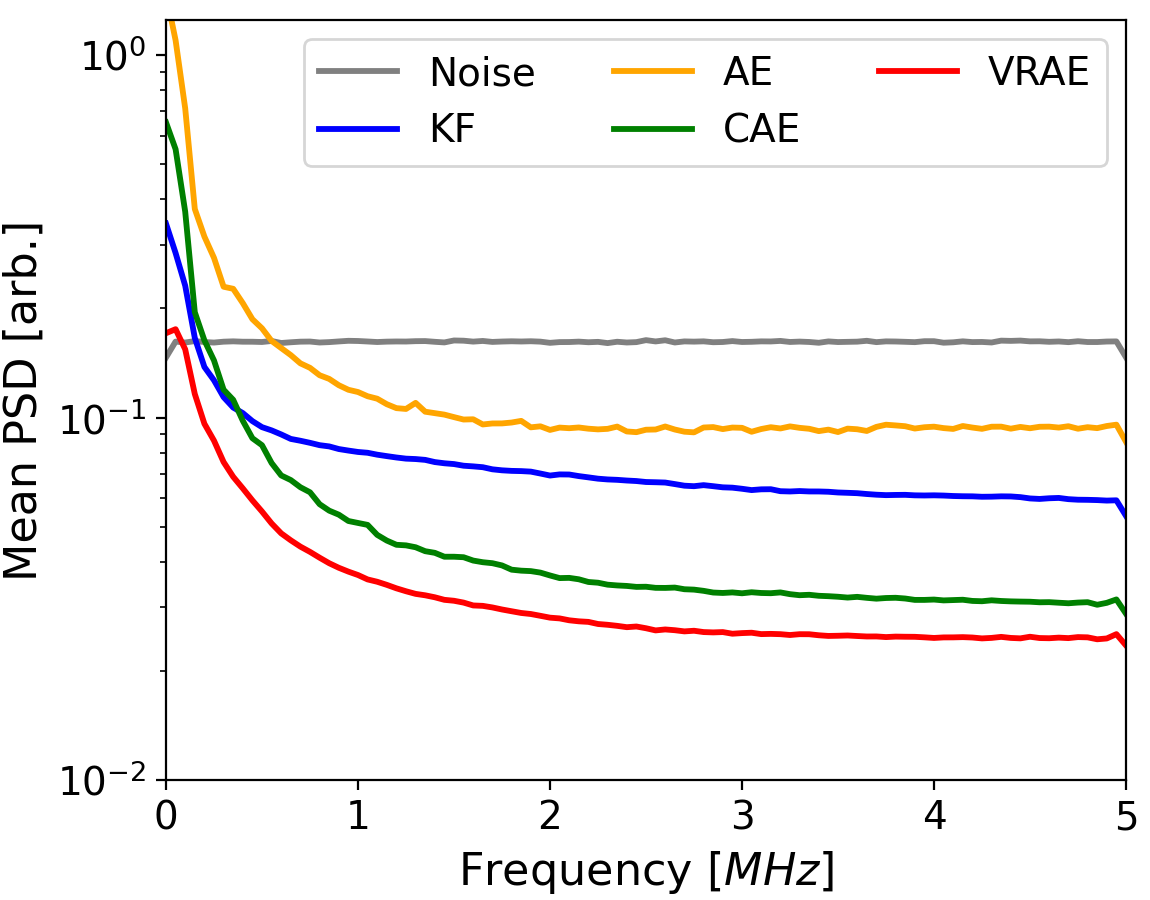}
    \caption{Mean noise power spectra for each model}
    \label{fig:mean_psds}
\end{wrapfigure}

To gain a better understanding of the overall performance of each model across the entire test set of data, we performed a thorough statistical analysis based on PSD and integrated noise outcomes. Figure \ref{fig:mean_psds} shows the mean PSD across the test set for each model. The averages for all models, with the exception of the simple AE, clearly beat the AWGN threshold from frequencies as low as $0.25$ MHz. The CAE and VRAE in particular achieve mean PSDs of less than 50\% and 16\%  of the AWGN intensity (respectively) across the majority of the available spectrum.

\begin{figure}[h!]
    \centering
    \includegraphics[width = 0.9\textwidth] {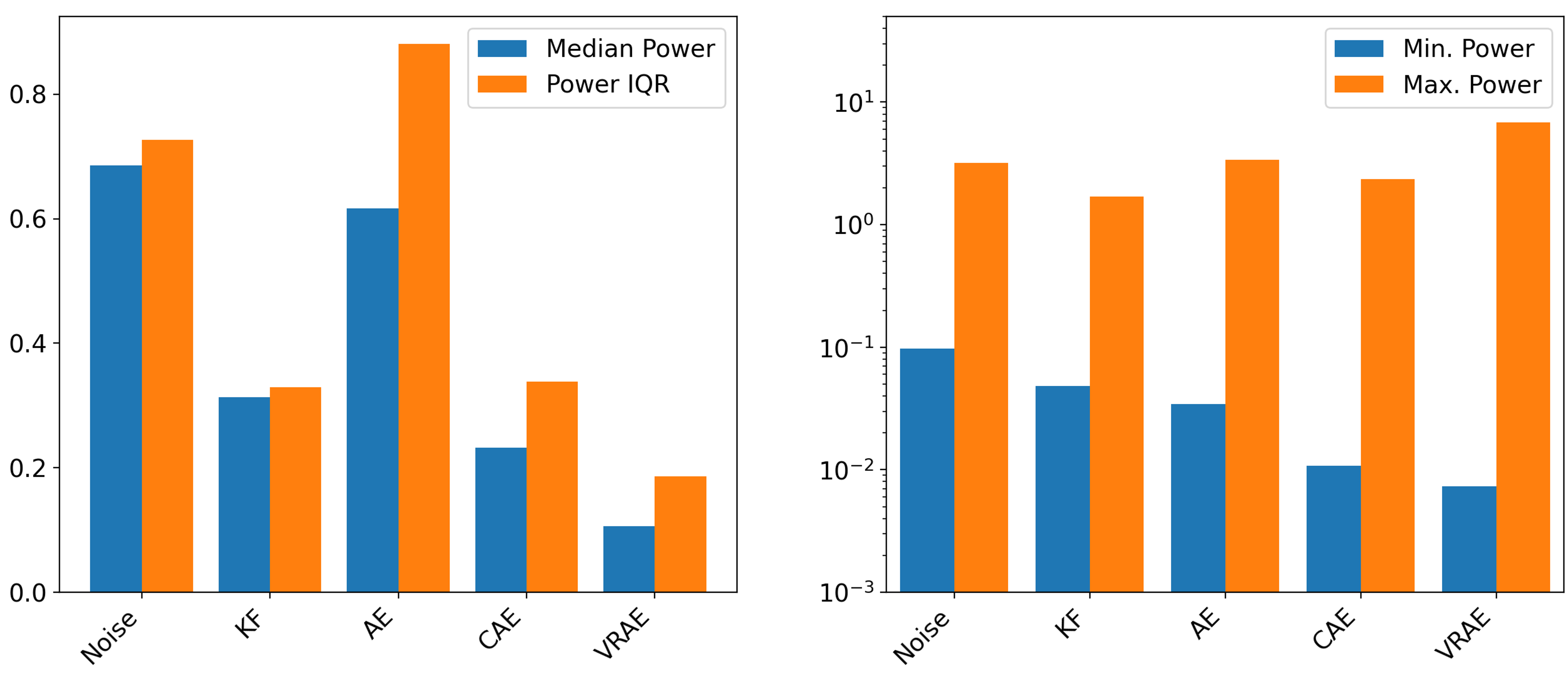}
    \caption{Total integrated noise medians and inter-quartile ranges (left) and extrema (right) over the test set for each method of noise reduction.}
    \label{fig:ipsd_bars}
\end{figure}

Figure \ref{fig:ipsd_bars} provides a view of average noise reduction across the spectrum in the form of total integrated noise medians and inter-quartile ranges (IQRs) (left) as well as the best and worst average performance in the form of total integrated noise minima and maxima (right). These results again demonstrate appreciable denoising by the KF by every metric, with substantially better performance achieved by the CAE and VRAE by every metric except for the maximum (worst-case) total integrated noise. Cases where the noise is increased by the AE, CAE, and VRAE are caused by the ML model not properly reconstructing the base waveform. In this instance the reconstruction error is manifested as \textbf{noise} in the power spectra. Here the KF has a clear benefit in that it reconstructs waveforms directly from the dynamics as opposed to a trained model. The feedforward autoencoder in particular is sensitive to these reconstruction error. 

\section{Conclusions}

We have explored four viable approaches for noise reduction in industrial RF signals: Kalman filters; feed-forward autoencoders; convolutional autoencoders; and variational autoencoders. Diverse training and test datasets were generated using an effective cavity circuit model thoroughly benchmarked against measured data.

Although each method exhibited noise reduction capabilities, the VRAE and CAE models achieved the best overall performance based on a variety of metrics. The KF model was found to have appreciable noise reduction capabilities despite its simplicity, especially given its lack of a need for training. Among our models, only the feed-forward AE failed to achieve robust noise reduction across the test dataset, though it proved capable of doing so for specific cases.

These results confirm the utility of ML methods for applications in RF signal processing and feedback controls in industrial settings. As such, a follow-up study of the potential for these models in related applications such as anomaly detection and data assimilation is now underway.

\acknowledgments

This material is based upon work supported by the U.S. Department of Energy, Office of Science, Office of Accelerator R\&D and Production Award Number DE-SC0023641.

\bibliographystyle{JHEP}
\bibliography{JINST_069P_0624}

\end{document}